\documentclass[twocolumn,superscriptaddress,prl,amsmath,amssymb,aps,showpacs]{revtex4}
\usepackage{graphicx}
\usepackage{bm}
\usepackage{endnotes}
\usepackage{amsmath,amssymb}
\newcommand{\rr}{{\bf r}}

\newcommand{\nc}{\newcommand}
\nc{\be}{\begin{equation}} \nc{\ee}{\end{equation}}
\nc{\bea}{\begin{eqnarray}} \nc{\eea}{\end{eqnarray}}
\nc{\bean}{\begin{eqnarray*}} \nc{\eean}{\end{eqnarray*}}
\nc{\mb}{\mbox} \nc{\rnc}{\renewcommand} \nc{\Bk}{\mathbf{k}}
\nc{\vp}{\mathbf{p}} \nc{\vn}{\mathbf{n}} \nc{\vq}{\mathbf{q}}
\nc{\Br}{\mathbf{r}} \nc{\vz}{\hat {\mathbf{z}}}
\nc{\vj}{\mathbf{j}} \nc{\vg}{\mathbf{g}}
\nc{\x}{\mathbf{x}} \nc{\A}{\mathbf{A}}
\nc{\va}{\mathbf{a}} \nc{\vs}{\mb{\boldmath$\sigma$}}
\nc{\vpi}{\mb{\boldmath$\pi$}} \nc{\nab}{\nabla} \nc{\X}{\sf x}
\nc{\Bsigma}{\mb{\boldmath$\sigma$}} \nc{\BB}{\mathbf{B}}

\begin{document}

\title{Absence  of skew scattering in two-dimensional systems:\\
Testing the origins of the anomalous Hall effect}
\author{Mario F. Borunda}
\affiliation{Department of Physics, Texas A\&M University,
College Station, TX 77843-4242, USA}
\author{Tamara S. Nunner}
\affiliation{Institut f\"ur Theoretische Physik, Freie Universit\"{a}t Berlin, Arnimallee 14, 14195 Berlin, Germany}
\author{Thomas L\"uck}
\affiliation{Institut f\"ur Theoretische Physik, Freie Universit\"{a}t Berlin, Arnimallee 14, 14195 Berlin, Germany}
\author{N. A. Sinitsyn}
\affiliation{CNLS/CCS-3, Los Alamos  National  Laboratory, Los Alamos, NM 87545, USA}
\affiliation{Department of Physics, Texas A\&M University,
College Station, TX 77843-4242, USA}
\author{Carsten Timm}
\affiliation{Department of Physics and Astronomy, University of Kansas, Lawrence, KS 66045, USA}
\author{J. Wunderlich}
\affiliation{Hitachi Cambridge Laboratory, Cambridge CB3 0HE, UK}
\author{T. Jungwirth}
\affiliation{Institute of Physics  ASCR, Cukrovarnick\'a 10, 162 53
Praha 6, Czech Republic }
\affiliation{School of Physics and Astronomy, University of Nottingham,
Nottingham NG7 2RD, UK}
\author{A. H. MacDonald}
\affiliation{Department of Physics, University of Texas at Austin,
Austin TX 78712-1081, USA}
\author{Jairo Sinova}
\affiliation{Department of Physics, Texas A\&M University,
College Station, TX 77843-4242, USA}

\date{February 12, 2007}

\begin{abstract}
We study the anomalous Hall conductivity in spin-polarized, asymmetrically
confined two-dimensional electron and hole systems, focusing on skew-scattering
contributions to the transport.   We find that  the skew scattering,
principally responsible for the extrinsic  contribution to the anomalous Hall
effect, vanishes for the two-dimensional electron system if both chiral Rashba
subbands are partially occupied, and vanishes always for the two-dimensional
hole gas studied here, regardless of the band filling.  Our prediction  can be
tested with the proposed coplanar two-dimensional electron/hole gas device and
can be used as a benchmark to understand  the crossover from the intrisic to
the extrinsic anomalous Hall effect.
\end{abstract}


\maketitle

\textit{Introduction}.---The observed Hall resistance of a magnetic  film
contains the ordinary Hall response to the external magnetic field and the
anomalous Hall response to the internal magnetization. Although the anomalous
Hall effect (AHE) has been used for decades as a basic characterization tool
for ferromagnets,  its origin is still being debated, also in the context of a
closely related novel phenomenon, the spin Hall effect
\cite{SHEtheory,Kato:2004_d,Wunderlich:2004_a,Sih:2005_a}. Three mechanisms
giving rise to AHE conductivity  have been identified: (1) an intrinsic
mechanism based solely on  the topological properties of the Bloch states
originating from the spin-orbit-coupled electronic structure 
\cite{Karplus:1954_a}, (2) a skew-scattering mechanism originating from the
asymmetry of the scattering rate  \cite{Smit:1955_a}, and (3) a side-jump
contribution, which semiclassically is viewed as  a side-step-type of
scattering and contributes to a net current perpendicular to the initial
momentum  \cite{Berger:1970_a}. 

Recent experimental and theoretical studies of tran\-si\-tion-me\-tal
ferromagnets and of less conventional systems, such as diluted magnetic
semiconductors, oxide and spinel ferromagnets, etc., have collected numerous
examples of the intrinsic  AHE  and of the transition to the extrinsic AHE
dominated by disorder scattering \cite{AHEmultiple}.  The unambiguous
determination of the origin of the AHE in these experimental systems is
hindered, in part, by their complex band structures, which has motivated
studies of simpler model Hamiltonians, such as the two-dimensional (2D) Rashba
and Dirac band models \cite{Dugaev:2005_a,
Sinitsyn:2005_a,Liu:2005_c,Inoue:2006_a,Onoda:2006_a}.  Attempts to describe
all the contributions to the AHE within the same framework have yielded
farraginous results, however. So far a rigorous connection of the more
intuitive semiclassical transport treatment with the  more systematic
diagramatic treatment, providing a clear-cut interpretation of the intrinsic,
skew, and side-jump AHE terms, has only been demonstrated for the Dirac
Hamiltonian model \cite{Sinitsyn:2006_b}.

In this Letter we calculate the transport coefficients in these two
complementary approaches for asymmetrically confined 2D electron and hole gases
in the presence of spin-independent disorder, finding perfect agreement. The
motivation for the study of these systems is threefold: First, they can be
represented by simple spin-orbit-coupled bands, which, similar to the Dirac
Hamiltonian model, allows us to unambiguously identify the individual AHE
contributions. Second, the extrinsic skew-scattering term vanishes for a
two-subband occupation in the case of the Rashba 2D electron gas and for
\textit{any} band occupation for the studied 2D hole gas.  This  provides a
clean  test of the intrinsic AHE mechanism and of the transition between the
intrinsic and skew-scattering-dominated AHE.  Finally,  we propose   a 2D
electron gas/2D hole gas coplanar magneto-optical device in which the unique
AHE phenomenology found in our theoretical models can be systematically
explored experimentally.

\textit{Model Hamiltonians}.---We study the following 2D model Hamiltonians:
\be
H = \frac{\hbar^2k^2}{2m} {\bm \sigma}_0 +i \alpha_n ({\bm \sigma}_+ k_-^n
  - {\bm \sigma}_- k_+^n) -h {\bm \sigma}_z+  V(\rr){\bm \sigma}_0
\label{ham}
\ee
with $m$ being the effective in-plane mass, ${\bm \sigma}_i$ the $2\times2$
Pauli matrices, $k_\pm = k_x\pm i k_y$, $h$ the exchange field, $\alpha_n$ the
spin-orbit coupling parameter, and $V(\rr)$ a spin-independent disorder
potential. The exponent $n=1$ ($3$) describes a 2D electron (hole) gas
\cite{Winkler:2003}. The eigenenergies of the clean system are $E_\pm = \hbar^2
k^2/2m \pm \sqrt{h^2 + (\alpha_n k^n)^2}$.  The eigenvectors in the clean
system take the form $\vert \Psi_{\Bk}^{\pm} \rangle = \exp(i \Bk \cdot \Br)
\vert u_{\Bk}^\pm \rangle$ with $\Bk = k\, (\cos \phi, \sin \phi)$ and
\be
\vert u_{\Bk}^\pm \rangle = \frac{1}{\sqrt{2\lambda}} \left(
\begin{array}{c}
\pm  i e^{-n i \phi} \sqrt{\lambda \pm h} \\
\sqrt{\lambda \mp h}\\ \end{array}
\right),
\label{basis}
\ee
where $\lambda = \sqrt{h^2 + (\alpha_n k^n)^2}$. We now define $k_\pm(E)$ as
the wave number for the $\pm$ band at a given energy $E$ and define
$\lambda_\pm\equiv\lambda(k_\pm)$. If $E$ is not specified, it is assumed to be
the Fermi energy. We consider the model of randomly located $\delta$-function
scatterers, $V({\bf r}) =\sum_i V_i\, \delta ({\bf r}-{\bf R}_i)$ with 
$\mathbf{R}_i$ random and disorder averages satisfying $\langle V_i
\rangle_{\mathrm{dis}} =0$, $\langle V_i^2 \rangle_{\mathrm{dis}} =V_0^2 \ne
0$,  and $\langle V_i^3 \rangle_{\mathrm{dis}} = V_1^3 \ne 0$.  This model is
different from the standard  white-noise disorder with $\langle |V^0_{{\bf
k'k}}|^2 \rangle_{\mathrm{dis}} = n_i V_0^2$, where $n_i$ is the impurity
concentration and other correlators are either zero or related to this
correlator by Wick's theorem. The deviation from white noise in our model is
quantified by $V_1\ne 0$, and is necessary to capture part of the
skew-scattering contribution to the AHE.

\textit{Semiclassical approach}.---We sketch here the semiclassical procedure
used in the calculation, for further details we refer  to Ref.\
\onlinecite{Sinitsyn:2006_b}. The multi-band Boltzmann equation  in a weak
electric field ${\bf E}$ is given by
\be
\label{e:Boltz}
\frac{\partial f_{l}}{\partial t} + e{\bf E}\cdot  {\bf v}_l
  \frac{d f_{l}}{d \epsilon} =
I[f]_{\mathrm{coll}}
\ee
where  $l=(\Bk, \mu$), $\mu=\pm$ is the subband index, and
$I[f]_{\mathrm{coll}} = - \sum_{\mu'} \int d^2 \Bk'/(2\pi)^2\: \omega_{ll'}
\left(f_{l} - f_{l'} \right)$ is the impurity collision integral.  The
distribution function $f_{l}$ is the sum of the  equilibrium function  and a
correction, $f_{l} = f^0_{l} + g_{l}$. The scattering rates $\omega_{ll'}$  are
related to the T-matrix elements through $\omega_{ll'} = 2 \pi/\hbar\, \vert
T_{l'l} \vert^2 \delta(\epsilon_{l'} - \epsilon_{l}),$ where $T_{l'l} = \langle
l' \vert V \vert \psi_l \rangle$, and $\vert \psi_l \rangle$ are eigenstates of
the complete Hamiltonian, and $|l\rangle\equiv |\Psi_{\mathbf{k}}^\mu\rangle$
of the disorder-free Hamiltonian.

\textit{Skew scattering}.---Skew scattering appears in the Boltzmann equation 
through the asymmetric part of  the scattering rate, i.e., $ \omega_{ll'}\neq
\omega_{l'l}$ \cite{Smit:1955_a}.  The scattering rates to second and third
order in disorder strength are given by
$\omega_{ll'}=\omega^{(2)}_{ll'}+\omega^{(3)}_ {ll'}+\cdots,$ where
$\omega^{(2)}_{ll'}=2\pi/\hbar\, \langle |V_{ll'}|^2 \rangle_{\mathrm{dis}}
\delta (\epsilon_{l} -\epsilon_{l'})$ is symmetric. Here $V_{l'l} = \langle l'
\vert V \vert l \rangle$. We break up the third-order contribution into
symmetric and antisymmetric parts. We ignore the first, since only the second
gives rise to  skew scattering. This antisymmetric term is given by
\cite{Sinitsyn:2006_b}
\be
\label{e:w3a} \omega_{ll'}^{(3a)} = -\frac{(2\pi)^2}{\hbar}  \sum \limits_{l''}
 \delta (\epsilon_{l}
 -\epsilon_{l''}) {\rm Im} \langle V_{ll'} V_{l'l''} V_{l''l}\rangle_{\mathrm{dis}} 
 \delta (\epsilon_{l} -\epsilon_{l'}) .
\ee
The solution of the Boltzmann equation (\ref{e:Boltz}) is found by first
looking at the deviation of the distribution function from equilibrium
\cite{Sinitsyn:2006_b},
\be
\label{e:dev_eqi}
g_l = - \frac{\partial f^0_\mu}{\partial \epsilon}\, e E \vert v_\mu \vert
  (A_\mu \cos\phi + B_\mu \sin\phi).
\ee
Assuming that the transverse conductivity is much smaller than the longitudinal
one ($A_\mu \gg B_\mu$) and substituting Eq.~(\ref{e:dev_eqi}) into
Eq.~(\ref{e:Boltz}) one finds $A_\mu = \tau_\mu^{\|}$ and $B_\mu =
{(\tau_\mu^{\|})^2}/{\tau_\mu^{\perp}}$, where 
\begin{eqnarray}
\label{e:time_v}
\frac{1}{\tau_\mu^{\|}} & = & \sum_{\mu'} \int \frac{d^2 \Bk'}{(2 \pi)^2}\,
  \omega_{ll'} \left[ 1 - \frac{\vert v_{l '}\vert}{\vert v_{l}\vert}
  \cos(\phi - \phi') \right],
\\
\label{e:time_p}
\frac{1}{\tau_\mu^{\perp}} & = & \sum_{\mu'} \int \frac{d^2 \Bk'}{(2 \pi)^2}\,
  \omega_{ll'}\, \frac{\vert v_{l '}\vert}{\vert v_{l}\vert}\, \sin(\phi - \phi') .
\end{eqnarray}
For symmetric Fermi surfaces, the skew-scattering contribution to the conductivity
tensor at zero temperature can now be expressed using the scattering times,
\be
\sigma_{x x}
= \frac{e^2}{4 \pi\hbar} \sum_\mu \tau_\mu^{\|} v_{F,\mu} k_\mu , \:
\sigma_{x y}^{\mathrm{\mathrm{skew}}} 
=\frac{e^2}{4 \pi\hbar} \sum_\mu \frac{(\tau_\mu^{\|})^2}{\tau_\mu^{\perp}} v_{F,\mu} k_\mu .
\label{e:zeroG_yx}
\ee
The calculation of $(\tau_\mu^{\|})^{-1}$ and $(\tau_\mu^{\perp})^{-1}$ uses the
matrix elements of Eq.~(\ref{e:w3a}). To simplify the notation we define
\be
\langle\mu\mu',\mu'\mu'',\mu''\mu \rangle \equiv \mbox{Im} \int_{0}^{2\pi} \!\!\!
  d\phi'' \langle u_{\Bk}^\mu \vert u_{\Bk'}^{\mu'} \rangle
  \langle u_{\Bk'}^{\mu'} \vert u_{\Bk''}^{\mu''} \rangle \langle u_{\Bk''}^{\mu''}
  \vert u_{\Bk}^\mu \rangle ,
\label{ee}
\ee
where  all momenta are taken on the Fermi surface. Note that in Eq.~(\ref{ee})
the magnitude of $\Bk''$ can be different from that of $\Bk'$ or $\Bk$ since
the Fermi momenta of different bands do not   coincide.

The matrix elements appearing in Eq.~(\ref{ee}) can be calculated directly from
the basis functions, yielding
\be
\langle \mu \mu',\mu'\mu'',\mu''\mu \rangle =
  - \frac{h \pi \alpha_n^2 k_\mu^n k_{\mu'}^n}{2 \lambda_\mu \lambda_{\mu'}
  \lambda_{\mu''}}\, \sin(n\phi - n\phi') , 
\label{e:mmmmmm}
\ee
from which we obtain
\be 
\omega_{ll'}^{(3a)} = -\frac{1}{\hbar}n_i  V_1^3
  \delta(\epsilon_l - \epsilon_{l'})\sum_{\mu''}  \nu^{\mu''}
  \langle \mu\mu',\mu'\mu'',\mu''\mu \rangle,
\label{e:w_3a--}
\ee
where $\nu^\pm$ is related to the density of states of each band at the Fermi
energy, $(\nu^\pm)^{-1}= {\hbar^2}/{m} \pm {n(\alpha_n k_\pm^{n-1})^2
}/{\lambda_\pm} $. The symmetric part of the scattering rates  to second order
in the disorder potential is  given by $\omega_{ll'}^{(2)} = 2 \pi/\hbar\, n_i
V_0^2  \vert \langle u_k^\mu \vert u_{k'}^{\mu '} \rangle \vert^2
\delta(\epsilon_l - \epsilon_{l'}).$ The relaxation times are found by
inserting this into Eq.~(\ref{e:time_v}) and Eq.~(\ref{e:w_3a--}) into
Eq.~(\ref{e:time_p}).   For $n=1$, i.e., for the 2D \textit{electron} gas, the
relaxation rates are then
\begin{eqnarray}
\frac{1}{\tau_\mu^{\|}} & = & \frac{1}{\hbar}\, n_i V_0^2 
  \bigg[ \frac{\nu^\mu}{\lambda_\mu} \left( \frac{h^2}{\lambda_\mu}
  + \frac{\alpha_1^2 k_\mu^2}{4 \lambda_+}+ \frac{\alpha_1^2 k_-^2}{4 \lambda_-}
  \right) \nonumber  \\
& & {}+ \frac{\nu^{\bar\mu}}{2} \left( 1 - \frac{h^2}{\lambda_- \lambda_+}
  \right) \bigg], \\
\label{tau-}
\frac{1}{\tau_\mu^{\perp}} & = & -\frac{n_i V_1^3  h \alpha_1^2\nu^\mu}
  {8\hbar\lambda_\mu}  \left(\frac{k_\mu^2}{\lambda_\mu}
  - \frac{k_{\bar\mu}^2}{\lambda_{\bar\mu}} \right) \left(\frac{\nu^\mu}{\lambda_\mu}
  - \frac{\nu^{\bar\mu}}{\lambda_{\bar\mu}}\right) ,\quad
\end{eqnarray}
where $\bar\mu\equiv-\mu$.  If both subbands are occupied, the last factor in
Eq.~(\ref{tau-}) vanishes and there is no skew-scattering contribution.  If
only the majority subband is occupied ($E_F < h$), $(\tau_\mu^{\perp})^{-1}$ is
non-zero and skew scattering contributes. For the skew-scattering Hall
conductivity and the longitudinal conductivity we obtain in this case
\begin{eqnarray}
\sigma_{x x} & = & \frac{e^2}{\pi \hbar n_i  V_0^2}
  \left(\frac{\lambda_- k_-}{\nu^-}\right)^{\!\!2} \frac{1}{3 h^2 + \lambda_-^2} , \\
\sigma_{xy}^{\mathrm{skew}} & = & -\frac{e^2  V_1^3 }{2 \pi \hbar n_i V_0^4 } \,
  \frac{h \lambda_- \alpha_1^2 k_-^4}{\nu_-(3h^2 + \lambda_-^2)^2}  .
\label{boltzskew}
\end{eqnarray}
If $n=3$, i.e., for the 2D \textit{hole} gas, we obtain
\begin{eqnarray}
\frac{1}{\tau_\mu^{\|}} & = & \frac{1}{\hbar}
  n_i  V_0^2  \left[ \frac{\nu^\mu}{2 \lambda_\mu^2} ( \lambda_\mu^2 + h^2 )
  + \frac{\nu^{\bar\mu} (\lambda_- \lambda_+ - h^2 )}{2 \lambda_- \lambda_+}
  \right] ,\quad \\
\frac{1}{\tau_\mu^{\perp}} & = & 0
\end{eqnarray}
and skew scattering vanishes irrespective of band filling.

\begin{figure}[t]
\includegraphics[width=.8\columnwidth]{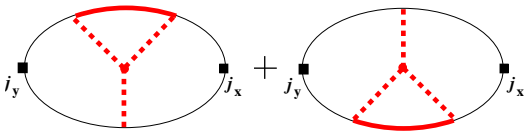}
\centering
\caption{(color online).
Diagramatic representation of the skew-scattering contribution to
$\sigma_{yx}$. Both current vertices, denoted by squares, are renormalized
by ladder vertex corrections.}  
\label{diagram1}
\end{figure}

\textit{Microscopic approach}.---Within the diagramatic
Kubo formalism the skew-scattering contribution to the
off-diagonal conductivity is obtained from the expression  
\begin{equation}
\sigma_{xy}^{I(a)}=\frac{e^2\hbar}{2\pi V} \sum_k 
{\rm Tr} \left[ v_x G_k^R(E_F) v_y G_k^A(E_F) \right] ,
\end{equation}
where the bare velocity vertex factors in the linear-in-$\mathbf{k}$ Rashba
model are given by
\begin{equation}
v_x=\frac{\hbar k_x}{m}{\bm \sigma}_0 -\frac{\alpha_1}{\hbar} {\bm \sigma}_y , \quad
v_y=\frac{\hbar k_y}{m}{\bm \sigma}_0 +\frac{\alpha_1}{\hbar} {\bm \sigma}_x.
\end{equation}
As shown in a previous study \cite{Sinitsyn:2006_b}, the skew-scattering
contribution  proportional to $V_1^3/(n_i V_0^4)$ corresponds to the diagrams
shown in  Fig.~\ref{diagram1}, where the current vertices $j_x, j_y$ on both
sides are the bare velocities $v_x, v_y$ renormalized by ladder vertex
corrections. Only the skew-scattering diagrams with a \textit{single}
third-order vertex, shown in Fig.~\ref{diagram1}, contribute to order
$V_1^3/(n_i V_0^4)$. All other terms from a ladder-type summation of
third-order vertices are smaller because they are either not of the order
$1/n_i$ or of higher order in $V_1/V_0$. The sum of the skew-scattering
vertices (i.e., the bold/red part of Fig.~\ref{diagram1}) gives
\begin{equation}
\frac{i}{4} n_i V_1^3 h \left( \frac{\nu_-}{\lambda_-} 
   - \frac{\nu_+}{\lambda_+} \right)
({\bm \sigma}_0 \otimes {\bm \sigma}_z - {\bm \sigma}_z \otimes {\bm \sigma}_0) \,.
\end{equation}
In the linear Rashba model we find $\nu_+/\lambda_+=\nu_-/\lambda_-$, implying
that skew scattering vanishes if both subbands are occupied. In the case that
only one subband is occupied the evaluation of Fig.~\ref{diagram1} to order
$V_1^3/(n_i V_0^4)$ yields exactly the same expression for $\sigma_{xy}^{\rm
skew}$ as in the semiclassical Eq.~(\ref{boltzskew}). The only effect of the
ladder vertex corrections is to renormalize each bare velocity by a factor of
$2(h^2+\lambda_-^2)/(3h^2+\lambda_-^2)$ which reduces to a factor of $1$ in the
limit of small $\alpha_1 k_F$ and to a factor of $2$ in the limit of small $h$.

For the 2D hole-gas model Hamiltonian (\ref{ham}) with $n=3$
the bare velocity vertex factors are
\begin{equation}
v_{x,y}=\frac{\hbar k_{x,y}}{m}{\bm \sigma_0}
  - \frac{6\alpha_3}{\hbar} k_x k_y {\bm \sigma}_{y,x}
  \pm \frac{3 \alpha_3}{\hbar} (k_x^2-k_y^2) {\bm \sigma}_{x,y} .
\end{equation}
Here the vertex corrections disappear because integrals of the type $\sum_k
G^R_k v_{x,y} G^A_k =0$ vanish. This implies the absence of skew scattering for
any subband filling \cite{Bernevig:2004_c},  consistent with the semiclassical
result. We note that the same consistency between semiclassical and microscopic
quantum theory calculations for the studied 2D models is also obtained for the
intrinsic and side-jump terms similar to the results in the graphene model
\cite{Sinitsyn:2006_b}; the longer details of those calculations will be shown
elsewhere and  are in general agreement with Ref.~
\onlinecite{Inoue:2006_a}.

The abscence of the skew scattering is akin but not  equivalent to the results
of spin-Hall-effect calculations in 2D systems \cite{Inoue:2004_a}. For the
Rashba 2D electron gas the disappearance of the DC spin Hall conductivity is
guaranteed by sum rules that relate the spin current to the dynamics of the
induced spin polarization \cite{Burkov:2004_a,Chalaev:2004_a}.  In the case of
a charge current no similar sum rule is known. As we have shown, the
skew-scattering contribution in fact becomes finite when the minority band is
depleated.  The vanishing of the Hall conductivity  in the Rashba  2D electron
gas for $E_F>h$ is attributed to the simplicity of the Hamiltonian. In
particular the relation $\nu_+/\lambda_+=\nu_-/\lambda_-$ does not hold
generally beyond the case of the linear-in-$\mathbf{k}$ Rashba coupling. The
abscence of skew scattering in the 2D hole system has a different origin: Due
to the cubic dependence of spin-orbit coupling on momentum, the matrix
elements, Eq.~(\ref{e:mmmmmm}), in the antisymmetric part of the collision term
behave like $\sin(3\phi-3\phi')$. Together with the $\sin(\phi-\phi')$
dependence of the velocity factor in Eq.~(\ref{e:time_p}), this makes the
integral over $\mathbf{k}'$ vanish.

Our results predict that the AHE in 2D electron and hole systems can be 
dominated by contributions independent of the impurity concentration, for which
the anomalous Hall resistance is $\propto \sigma_{xx}^{-2}$.  We also predict
that in the Rashba 2D electron gas with only one subband occupied the extrinsic
skew-scattering contribution, leading to anomalous Hall resistance proportional
to $\sigma_{xx}^{-1}$, is non-zero. Note that  this term has not been
identified in previous works that considered only white-noise disorder
\cite{Dugaev:2005_a, Sinitsyn:2005_a,Liu:2005_c,Inoue:2006_a,Onoda:2006_a}.
Since its corresponding conductivity contribution is inversely proportional to
the impurity concentration, the skew-scattering mechanism can  dominate in
clean samples.

\begin{figure}[t]
\includegraphics[width=\columnwidth]{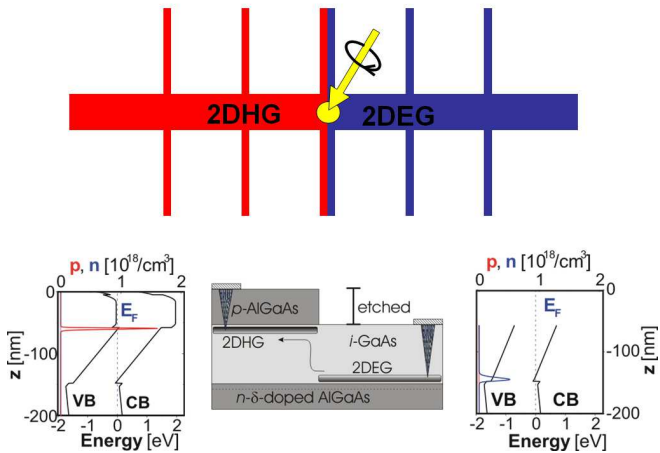}
\centering
\caption{(color online). Top panel: Top-view schematics of the Hall bar with
coplanar 2D hole and electron gases. Spin-polarized carriers are generated by
shining circularly polarized light on the p-n junction. Center bottom panel:
Cross section of the heterostructure containing p-type and n-type AlGaAs/GaAs
single junctions. The left band diagram corresponds to the unetched part of the
wafer with the 2D hole gas, the right band diagram shows the 2D electron gas in
the etched section of the wafer.}  
\label{experiment}
\end{figure}

\textit{Proposed experimental setup}.---The unique phe\-no\-me\-no\-lo\-gy of
the AHE in the studied 2D systems, in particular  the sudden disappearance of
skew scattering when the Fermi level crosses the depletion point of the
minority 2D Rashba band, represents an opportunity for a clean test of the
presence of intrinsic and extrinsic sources of the AHE and of the transition
between these two regimes.   In the absence of 2D ferromagnetic system with
Rashba like spin-orbit interatciton, we proposed an experimental setup for this
test as shown in Fig.~\ref{experiment}. The device is based on a AlGaAs/GaAs
heterostructure containing a coplanar 2D hole gas/2D electron gas p-n junction.
The cross section of the heterostructure and corresponding band diagrams are
shown in the lower panels of Fig.~\ref{experiment} (for more details see
Ref.~\onlinecite{Wunderlich:2004_a}). Under a forward bias the junction was
successfully utilized as a light-emitting-diode spin detector for the spin Hall
effect \cite{Wunderlich:2004_a}. Here we propose to operate the junction in the
reverse-bias mode, while  shining monochromatic, circularly polarized light of
tuneable wavelength on  the p-n junction. The photogenerated spin-polarized
holes and electrons will propagate in opposite directions through the
respective 2D hole and electron channels. The longitudinal voltage and the
generated anomalous Hall voltage can be detected by the successive sets of Hall
probes, as shown in the upper panel of Fig.~\ref{experiment}. For the 2D
electron gas the macroscopic spin diffusion length allows to use standard
lithography for defining the Hall probes.  Surface or back gates in close
proximity to the 2D electron system can be used to modify the effective 2D
confinements, carrier density, and spin-orbit coupling in order to control the
transition between the intrinsic and extrinsic AHE regimes. The exploration of
the AHE in the 2D hole gas is more challenging due to the expected sub-micron
spin diffusion length in this system but may still be feasible in the proposed
experimental setup.

\begin{acknowledgments}
Fruitful discussions with S. Onoda and N. Nagaosa are gratefully
acknowledged. This work was supported by ONR under Grant No.\
ONR-N000140610122, by the NSF under Grants No.\ DMR-0547875 and No.\
PHY99-07949, by the SRC-NRI (SWAN), by EU Grant IST-015728, by EPSRC Grant
GR/S81407/01, by GACR and AVCR Grants 202/05/0575, FON/06/E002, AV0Z1010052,
and LC510, by the DOE under Grant No.\ DE-AC52-06NA25396, and by the Univ. of
Kansas General Research Fund allocation No.\ 2302015. Jairo Sinova is a
Cottrell Scholar of Research Corporation.
\end{acknowledgments}


\begin{thebibliography}{24}
\expandafter\ifx\csname natexlab\endcsname\relax\def\natexlab#1{#1}\fi
\expandafter\ifx\csname bibnamefont\endcsname\relax
  \def\bibnamefont#1{#1}\fi
\expandafter\ifx\csname bibfnamefont\endcsname\relax
  \def\bibfnamefont#1{#1}\fi
\expandafter\ifx\csname citenamefont\endcsname\relax
  \def\citenamefont#1{#1}\fi
\expandafter\ifx\csname url\endcsname\relax
  \def\url#1{\texttt{#1}}\fi
\expandafter\ifx\csname urlprefix\endcsname\relax\def\urlprefix{URL }\fi
\providecommand{\bibinfo}[2]{#2}
\providecommand{\eprint}[2][]{\url{#2}}

\bibitem{SHEtheory}
\bibinfo{author}{\bibfnamefont{M.~I.} \bibnamefont{Dyakonov}}
  \bibnamefont{and}  \bibinfo{author}{\bibfnamefont{V.~I.} \bibnamefont{Perel}},
  \bibinfo{journal}{JETP} \bibinfo{pages}{467} (\bibinfo{year}{1971});
  \bibinfo{author}{\bibfnamefont{S.}~\bibnamefont{Murakami}},
  \bibinfo{author}{\bibfnamefont{N.}~\bibnamefont{Nagaosa}}, \bibnamefont{and}
  \bibinfo{author}{\bibfnamefont{S.-C.} \bibnamefont{Zhang}},
  \bibinfo{journal}{Science} \textbf{\bibinfo{volume}{301}},
  \bibinfo{pages}{1348} (\bibinfo{year}{2003});
  \bibinfo{author}{\bibfnamefont{J.}~\bibnamefont{Sinova}} {\it et al.},
  \bibinfo{journal}{Phys. Rev. Lett.} \textbf{\bibinfo{volume}{92}},
  \bibinfo{pages}{126603} (\bibinfo{year}{2004}).

\bibitem{Kato:2004_d}
\bibinfo{author}{\bibfnamefont{Y.~K.} \bibnamefont{Kato}} {\it et al.},
  \bibinfo{journal}{Science}  \textbf{\bibinfo{volume}{306}},
  \bibinfo{pages}{1910} (\bibinfo{year}{2004}).

\bibitem{Wunderlich:2004_a}
\bibinfo{author}{\bibfnamefont{J.}~\bibnamefont{Wunderlich}} {\it et al.},
  \bibinfo{journal}{Phys. Rev. Lett.} \textbf{\bibinfo{volume}{94}},
  \bibinfo{pages}{047204} (\bibinfo{year}{2005}).

\bibitem{Sih:2005_a}
\bibinfo{author}{\bibfnamefont{V.}~\bibnamefont{Sih}} {\it et al.},
  \bibinfo{journal}{Nature Physics} \textbf{\bibinfo{volume}{1}},
  \bibinfo{pages}{31} (\bibinfo{year}{2005}).

\bibitem{Karplus:1954_a}
\bibinfo{author}{\bibfnamefont{R.}~\bibnamefont{Karplus}} \bibnamefont{and}
  \bibinfo{author}{\bibfnamefont{J.~M.} \bibnamefont{Luttinger}}, 
  \bibinfo{journal}{Phys. Rev.} \textbf{\bibinfo{volume}{95}},
  \bibinfo{pages}{1154} (\bibinfo{year}{1954}).

\bibitem{Smit:1955_a}
\bibinfo{author}{\bibfnamefont{J.}~\bibnamefont{Smit}},
  \bibinfo{journal}{Physica} \textbf{\bibinfo{volume}{21}}, \bibinfo{pages}{877}
  (\bibinfo{year}{1955}).

\bibitem{Berger:1970_a}
\bibinfo{author}{\bibfnamefont{L.}~\bibnamefont{Berger}},  \bibinfo{journal}{Phys. Rev.}
  \textbf{\bibinfo{volume}{B 2}}, \bibinfo{pages}{4559} (\bibinfo{year}{1970}).

\bibitem{AHEmultiple}
\bibinfo{author}{\bibfnamefont{J.}~\bibnamefont{Banhart}} \bibnamefont{and} 
  \bibinfo{author}{\bibfnamefont{H.}~\bibnamefont{Ebert}},
  \bibinfo{journal}{Europhys. Lett.} \textbf{\bibinfo{volume}{32}},
  \bibinfo{pages}{517} (\bibinfo{year}{1995});
  \bibinfo{author}{\bibfnamefont{J.}~\bibnamefont{Ye}} {\it et al.},
  \bibinfo{journal}{Phys. Rev. Lett.} \textbf{\bibinfo{volume}{83}},
  \bibinfo{pages}{3737} (\bibinfo{year}{1999});
  \bibinfo{author}{\bibfnamefont{T.}~\bibnamefont{Jungwirth}},
  \bibinfo{author}{\bibfnamefont{Q.}~\bibnamefont{Niu}}, \bibnamefont{and}
  \bibinfo{author}{\bibfnamefont{A.~H.} \bibnamefont{MacDonald}},  {\it ibid.}
  \textbf{\bibinfo{volume}{88}},  \bibinfo{pages}{207208} (\bibinfo{year}{2002});
  \bibinfo{author}{\bibfnamefont{Y.}~\bibnamefont{Yao}} {\it et al.}, {\it ibid.}
  \textbf{\bibinfo{volume}{92}},  \bibinfo{pages}{037204} (\bibinfo{year}{2004});
  \bibinfo{author}{\bibfnamefont{Y.}~\bibnamefont{Taguchi}} {\it et al.},
  \bibinfo{journal}{Science} \textbf{\bibinfo{volume}{291}}, \bibinfo{pages}{5513}
  (\bibinfo{year}{2001});
  \bibinfo{author}{\bibfnamefont{W.-L.} \bibnamefont{Lee}} {\it et al.}, {\it ibid.}
  \textbf{\bibinfo{volume}{303}},  \bibinfo{pages}{1647} (\bibinfo{year}{2004});
  \bibinfo{author}{\bibfnamefont{J.}~\bibnamefont{{K\"{o}tzler}}}  \bibnamefont{and}
  \bibinfo{author}{\bibfnamefont{W.}~\bibnamefont{Gil}},  \bibinfo{journal}{Phys. Rev.}
  \textbf{\bibinfo{volume}{B 72}},  \bibinfo{pages}{060412(R)} (\bibinfo{year}{2005});
  \bibinfo{author}{\bibfnamefont{B.~C.} \bibnamefont{Sales}} {\it et al.},  {\it ibid.}
  \textbf{\bibinfo{volume}{ 73}},  \bibinfo{pages}{224435} (\bibinfo{year}{2006});
  \bibinfo{author}{\bibfnamefont{C.}~\bibnamefont{Zeng}} {\it et al.},
  \bibinfo{journal}{Phys. Rev. Lett.} \textbf{\bibinfo{volume}{96}},
  \bibinfo{pages}{037204} (\bibinfo{year}{2006});
  \bibinfo{author}{\bibfnamefont{S.~H.} \bibnamefont{Chun}} {\it et al.},
  {\it ibid.}  \textbf{\bibinfo{volume}{98}}, \bibinfo{pages}{026601}  (\bibinfo{year}{2007});
  J. Cumings {\it et al.}, {\it ibid.} {\bf 96}, 196404 (2006);
  \bibinfo{author}{\bibfnamefont{T.}~\bibnamefont{Miyasato}} {\it et al.},
  \eprint{cond-mat/0610324}.

\bibitem[{\citenamefont{Dugaev et~al.}(2005)\citenamefont{Dugaev, Bruno,
  Taillefumier, Canals, and Lacroix}}]{Dugaev:2005_a}
\bibinfo{author}{\bibfnamefont{V.~K.} \bibnamefont{Dugaev}} {\it et al.},
  \bibinfo{journal}{Phys. Rev.} \textbf{\bibinfo{volume}{B 71}},
  \bibinfo{pages}{224423} (\bibinfo{year}{2005}).

\bibitem[{\citenamefont{Sinitsyn et~al.}(2005)\citenamefont{Sinitsyn, Niu,
  Sinova, and Nomura}}]{Sinitsyn:2005_a}
\bibinfo{author}{\bibfnamefont{N.~A.} \bibnamefont{Sinitsyn}} {\it et al.},
  \bibinfo{journal}{Phys. Rev.} \textbf{\bibinfo{volume}{B 72}},
  \bibinfo{pages}{045346} (\bibinfo{year}{2005}).

\bibitem[{\citenamefont{Liu and Lei}(2005)}]{Liu:2005_c}
\bibinfo{author}{\bibfnamefont{S.~Y.} \bibnamefont{Liu}} \bibnamefont{and}
  \bibinfo{author}{\bibfnamefont{X.~L.} \bibnamefont{Lei}},
  \bibinfo{journal}{Phys. Rev.} \textbf{\bibinfo{volume}{B 72}},
  \bibinfo{pages}{195329} (\bibinfo{year}{2005}).

\bibitem[{\citenamefont{ichiro Inoue et~al.}(2006)\citenamefont{ichiro Inoue,
  Kato, Ishikawa, Itoh, Bauer, and Molenkamp}}]{Inoue:2006_a}
\bibinfo{author}{\bibfnamefont{J.}~\bibnamefont{Inoue}} {\it et al.},
\bibinfo{journal}{Phys. Rev. Lett.}
  \textbf{\bibinfo{volume}{97}}, \bibinfo{pages}{046604}
  (\bibinfo{year}{2006}).

\bibitem[{\citenamefont{Onoda et~al.}(2006)\citenamefont{Onoda, Sugimoto, and
  Nagaosa}}]{Onoda:2006_a}
\bibinfo{author}{\bibfnamefont{S.}~\bibnamefont{Onoda}},
  \bibinfo{author}{\bibfnamefont{N.}~\bibnamefont{Sugimoto}}, \bibnamefont{and}
  \bibinfo{author}{\bibfnamefont{N.}~\bibnamefont{Nagaosa}},
 Phys. Rev. Lett. {\bf 97}, 126602 (2006).

\bibitem[{\citenamefont{Sinitsyn et~al.}(2006)\citenamefont{Sinitsyn,
  MacDonald, Jungwirth, Dugaev, and Sinova}}]{Sinitsyn:2006_b}
\bibinfo{author}{\bibfnamefont{N.~A.} \bibnamefont{Sinitsyn}} {\it et al.},
  Phys. Rev. B {\bf 75}, 045315 (\bibinfo{year}{2007}).

\bibitem{Winkler:2003} R. Winkler, {\it Spin-Orbit Coupling Effects in Two-
Dimensional Electron and Hole Systems}, Springer Tracts in Modern Physics, vol.\ 191 (Springer,
Berlin, 2003).

\bibitem[{\citenamefont{Bernevig and Zhang}(2005)}]{Bernevig:2004_c}
\bibinfo{author}{\bibfnamefont{B.~A.} \bibnamefont{Bernevig}} \bibnamefont{and}
  \bibinfo{author}{\bibfnamefont{S.-C.} \bibnamefont{Zhang}},
  \bibinfo{journal}{Phys. Rev. Lett.} \textbf{\bibinfo{volume}{95}},
  \bibinfo{pages}{016801} (\bibinfo{year}{2005}).

\bibitem[{\citenamefont{Inoue et~al.}(2004)\citenamefont{Inoue, Bauer, and
  Molenkamp}}]{Inoue:2004_a}
\bibinfo{author}{\bibfnamefont{J.}~\bibnamefont{Inoue}},
  \bibinfo{author}{\bibfnamefont{G.~E.~W.} \bibnamefont{Bauer}},
  \bibnamefont{and} \bibinfo{author}{\bibfnamefont{L.~W.}
  \bibnamefont{Molenkamp}}, \bibinfo{journal}{Phys. Rev.}
  \textbf{\bibinfo{volume}{B 70}}, \bibinfo{pages}{041303}
  (\bibinfo{year}{2004}), \eprint{cond-mat/0402442}.

\bibitem{Burkov:2004_a}
\bibinfo{author}{\bibfnamefont{A.~A.} \bibnamefont{Burkov}},
  \bibinfo{author}{\bibfnamefont{A.~S.} \bibnamefont{Nu\~{n}ez}}, \bibnamefont{and}
  \bibinfo{author}{\bibfnamefont{A.~H.} \bibnamefont{MacDonald}},
  \bibinfo{journal}{Phys. Rev.} \textbf{\bibinfo{volume}{B 70}},
  \bibinfo{pages}{155308} (\bibinfo{year}{2004}).

\bibitem[{\citenamefont{Chalaev and Loss}(2004)}]{Chalaev:2004_a}
\bibinfo{author}{\bibfnamefont{O.}~\bibnamefont{Chalaev}} \bibnamefont{and}
  \bibinfo{author}{\bibfnamefont{D.}~\bibnamefont{Loss}},
  \bibinfo{journal}{Phys. Rev.} \textbf{\bibinfo{volume}{B 71}},
  \bibinfo{pages}{245318} (\bibinfo{year}{2004}).


\end{thebibliography}
\end{document}